\documentclass[10pt]{article}
\usepackage{a4wide}                      
\usepackage{epsf}                        
\usepackage{latexsym}                    
\usepackage{amsfonts}                    
\usepackage{amssymb}                     
\usepackage{slashed}                     

\newcommand{\sect}[1]{\setcounter{equation}{0}\section{#1}}
\renewcommand{\theequation}{\arabic{section}.\arabic{equation}}

\def\be{\begin{equation}}
\def\ee{\end{equation}}
\def\bea{\begin{eqnarray}}
\def\eea{\end{eqnarray}}
\def\nn{\nonumber \\}
\def\vsp#1{\vspace{#1}}
\def\hsp#1{\hspace{#1}}

\def\part{\partial}
\def\tfrac#1#2{{\textstyle{\frac{#1}{#2}}}}
\def\half{\tfrac{1}{2}}

\newcommand{\ft}[2]{{\textstyle\frac{#1}{#2}}}
\newcommand{\for}[1]{(\ref{#1})}
\def\rmi{{\,\rm i\,}}

\def\mn{{\mu\nu}}
\def\mnr{{\mu\nu\rho}}

\def\Ortin{Ort{\'\i}n}

\def\makeatletter{\catcode`\@=11}
\makeatletter
\def\mathbox#1{\hbox{$\m@th#1$}}%
\def\math@ccstyles#1#2#3#4#5#6#7{{\leavevmode
      \setbox0\mathbox{#6#7}%
      \setbox2\mathbox{#4#5}%
      \dimen@ #3%
      \baselineskip\z@\lineskiplimit#1\lineskip\z@
      \vbox{\ialign{##\crcr
             \hfil \kern #2\box2 \hfil\crcr
             \noalign{\kern\dimen@}%
             \hfil\box0\hfil\crcr}}}}
\def\mathaccstyles{\math@ccstyles\maxdimen}
\def\maththroughstyles{\math@ccstyles{-\maxdimen}}
\def\unity%
 {\maththroughstyles{.45\ht0}\z@\displaystyle {\mathchar"006C}\displaystyle 1}
\begin{document}

~
\vsp{1.5cm}

\rightline{KUL-TF/03-03}
\rightline{hep-th/0301184}
\rightline{January 2003}
\vspace{3truecm}

\centerline{\Large \bf  Probes in fluxbranes and supersymmetry}
\vspace{.6truecm}
\centerline{\Large \bf breaking through Hodge-duality}
\vspace{2truecm}

\centerline{
    {\bf Jos Gheerardyn\footnote{Email: {\tt jos.gheerardyn@fys.kuleuven.ac.be}}  \ 
 and\ \ Bert Janssen\footnote{Email: {\tt bert.janssen@fys.kuleuven.ac.be}}  
}
                                                            }

\vspace{1cm}
\centerline{{\it Instituut voor Theoretische Fysica, }}
\centerline{{\it Katholieke Universiteit Leuven,}}  
\centerline{{\it Celestijnenlaan 200D,  B-3001 Leuven, Belgium}}

\vspace{3truecm}

\centerline{\bf ABSTRACT}
\vspace{.5truecm}
\noindent
In this letter we consider what happens to an M2-brane probe when Minkowski space is dimensionally reduced to
a fluxbrane solution of IIA supergravity. Given that fluxbrane reductions generally break supersymmetry, we look at how supersymmetry is realised on the D2-brane probe after dualisation. We also 
show how to extend this to more general non-linear sigma models.

\newpage
\sect{Introduction}

It is standard lore that, at the level of the low energy effective action, dimensional 
reduction and T-duality can break supersymmetry \cite{Bakas, BKO}: two supergravity 
solutions related via T-duality or via dimensional reduction do not have in general the 
same number of supersymmetries. Probably the best-known example is the one obtained by 
dimensionally reducing Minkowski space, written in spherical coordinates, over an angular 
Killing vector $\part_\varphi$. Minkowski space is of course maximally supersymmetric, but 
the obtained solution has no surviving supersymmetries. Many other examples were given in
\cite{Bakas, BKO}. The breaking of supersymmetry in these cases is due to the fact that, 
although the (bosonic) solution admits an isometry, the Killing spinors are not invariant 
under its action. In order for supersymmetry to be preserved in the dualisation or 
dimensional reduction of supergravity solutions, it is therefore necessary that the 
Killing spinors are independent of the isometry direction that is considered \cite{BKO}.

This is of course an artefact of the fact we are only dealing with the low energy 
effective action. If one would take in account the full string spectrum with all 
higher-order Kaluza-Klein modes, then it is clear that supersymmetry is preserved, since 
dimensional reduction and dualisations are merely rewritings and canonical transformations 
\cite{AAL}. Indeed, it has been shown \cite{BS, Hassan, AAB, Sfetsos} that in the 
dualisation local realisations of supersymmetry can be replaced by non-local realisations, 
due to non-local world-sheet effects, but this will never break supersymmetry at the level 
of the conformal field theory. Only while looking at the lowest-mode approximation, 
supersymmetry can be lost in the truncation.

There are some well known examples where specific use is made of dimensional reduction
along non-trivial Killing vectors (involving both translations and rotations) in order to 
obtain new solutions. In \cite{GW, GM, DGGiH} it was realised that a dilatonic version of 
the Melvin universe \cite{Melvin} can be obtained from dimensional reduction of flat 
space-time, if the higher-dimensional solution has some non-trivial identifications 
\cite{DGKT, DGGH, DGGH2}. The lower-dimensional solution describes a fluxtube, with the 
vector potential being the Kaluza-Klein vector from the reduction. In 
\cite{RT} a ten-dimensional version of this fluxtube (F7-brane in modern terminology) was 
used to study the Green-Schwarz string action in backgrounds with RR-fields, in 
\cite{CG} to point out a duality between Type IIA and Type 0A, and in \cite{CHC, emparan2, 
BSa} to construct supergravity solutions of D-branes and F-strings undergoing the 
dielectric effect \cite{myers}. These solutions 
are in general non-sypersymmetric due to the dependence of the Killing spinors on the 
compactified coordinate. 

However, recently in \cite{GS, Uranga, RT2} it was found that in some specific cases, 
involving several carefully chosen (magnetic) parameters, some amount of supersymmetry can 
be preserved in the dimensional reduction. In \cite{FS} a systematic study of these 
cases was made: it turned out that dimensional reduction of flat space along a Killing 
vector consisting of a translation and a rotation will preserve some fraction of the 
supersymmetry if the rotation lays in the isotropy algebra of the Killing spinor, and 
these isotropy algebras were classified. In this way, the Killing vector will leave some 
Killing spinors invariant and the amount of preserved supersymmetry is given by the number 
of Killing spinor components that are independent of the coordinate associated to the 
isometry direction. In this way several supersymmetric fluxbranes have been constructed. 

In this letter, we will use probe techniques to study the (partial) breaking of 
supersymmetry in the construction of these new fluxbrane solutions.  More 
specifically, we will study a simple toy model and see what happens with an M2-brane probe 
in eleven-dimensional Minkowski space, when the space is dimensionally reduced to a 
fluxbrane solution. The reduction is transverse to the probe such that the M2-brane turns 
into a D2-brane probe in the fluxbrane background. Given that the lower-dimensional 
background has less supersymmetry than the original Minkowski space, it is to be expected 
that also the world-volume theory living on the D2-brane probe will be less supersymmetric 
then the one living on the M2-brane at the level of the non-linear sigma model. Indeed, 
it is clear from \cite{Bakas, BKO} that (some) supersymmetry will be broken due to 
explicit dependence of the Killing spinor on the isometry direction. We will show that, 
as in \cite{FS}, with specific choices of the rotational parameters, the world-volume 
theory on the D2-brane will preserve a certain amount of supersymmetry. The difference 
with the construction of \cite{FS} lays in the fact that here, in the world-volume theory, 
it is not the dimensional reduction directly that is responsible for the 
partial supersymmetry breaking, but the Hodge-dualisation of one of the embedding scalars 
of the M2-brane into the Born-Infeld vector of the D2-brane \cite{Towns}. The amount of 
supersymmetry on the world-volume after dualisation will be given by the number of 
supersymmetry parameters that are independent of the dualised coordinate.

This paper is organised as follows: in Section 2 we revise shortly the construction of 
supersymmetric fluxbranes of \cite{FS} and discuss the conditions for supersymmetry
to be preserved in these solutions. In Section 3 we consider the world-volume theories
that live on the M2- and D2-brane probes, performing a dimensional reduction of the type
done in \cite{FS}, we see how the Hodge-dualisation of the isometric scalar has to be 
performed for the M2-brane world-volume action to be reduced to the D2-brane action, first 
in the bosonic case and then in the supersymmetric case. By looking at the supersymmetry 
algebra and the invariance of the action in the reduced theory, we show that the same 
conditions for supersymmetry preservation hold for the world-volume fields as for the 
backgrounds. In Section 4 we argue that this procedure is quite general and that the 
Hodge-dualisation of a scalar in any non-linear sigma-model is compatible with (some part of) the supersymmetry-algebra, as
long as the Killing vector associated to the dualised scalar commutes with some subgroup 
of the super-Poincar\'e algebra. Finally, our conventions on the spinors of the different 
groups appearing in this letter are given in the Appendix.

\sect{Reductions of Minkowski space}

It is well known that the ten-dimensional dilatonic Melvin universe \cite{GW, GM, DGGiH, 
RT, CG} can be obtained via a dimensional reduction of eleven-dimensional Minkowski 
space with some non-trivial identifications \cite{DGKT, DGGH, DGGH2}. After 
a reduction over a Killing vector $\xi$ consisting of a combination of a 
translation and a rotation $B$, 
\begin{equation}
\xi=\partial_z + B_i{}^j x^i \partial_j\; ,
\label{xi}
\end{equation}
a so-called fluxbrane is obtained, where the Kaluza-Klein vector gives rise 
to a flux tube in the lower-dimensional solution. It has been pointed out in 
\cite{GS, Uranga, RT2, FS} that for specific choices of the rotation $B$, the 
obtained fluxbranes could preserve some amount of supersymmetry. In this 
Section we will follow closely the construction presented in \cite{FS}. 

The first step in the dimensional reduction is choosing a coordinate system 
of Minkowski space adapted to the Killing vector, such that $\xi=\partial_z$ 
We start from the standard coordinates on eleven-dimensional Minkowski space. 
Firstly, we write the Killing vector \for{xi} as
\begin{equation}
\xi=U^{-1}\partial_z U,  \hsp{1.5cm} 
 U=e^{z B}\; .
\end{equation}
The coordinate system adapted to the Killing vector $\xi$ is then defined by 
\begin{equation}\label{chco}
\vec{y} = U^{-1} \vec{x}\; ,
\end{equation} 
since in these coordinates $\{y^i,z\}$ we have $\xi \vec{y}=0.$ In these coordinates 
Minkowski space is given by\footnote{The Greek indices $\mu$ run from 0 to 2 
and the Latin indices $i$ from 1 to 7. This specific splitting of the 
eleven-dimensional indices is motivated by that fact that we will later put an 
M2- and D2-brane probe in the $x^\mu$-directions. We therefore restricted the 
rotation on the Killing vector deliberately to the subspace transverse to the 
$x^\mu$s. This choice will of course limit the number of possible 
supersymmetric solutions.}
\begin{eqnarray}
d{\hat s}^2 &=&\eta_{\mu \nu} dx^\mu dx^\nu 
        + \Big( \delta_{ij} - \Lambda^{-1}  B_{ik} y^k B_{jm} y^m\Big)dy^idy^j
    +\Lambda\Big(dz- \Lambda^{-1} B_{ij} y^i dy^j \Big)^2, \\[.3cm]
\Lambda &=& 1 - y^i B_{ij}B_{jk} y^k. \nonumber
\end{eqnarray}
Using the standard Kaluza-Klein 
Ansatz, the corresponding Type IIA solution is given by
\begin{eqnarray}
ds^2&=&\Lambda^{\frac12}\eta_{\mu \nu} dx^\mu dx^\nu 
         +\Lambda^{\frac12} 
          \Bigl(\delta_{ij} +\Lambda^{-1} B_{ik}y^k B_{jl} y^l \Bigr)dy^i dy^j
\nonumber\\
\phi&=&\ft34 \ln \Lambda
\label{solIIA}\\
C_i&=& \Lambda^{-1} B_{ij}y^j\; . 
\nonumber
\end{eqnarray}

It is interesting to know for what choices of $B$ the solution (\ref{solIIA}) 
preserves some supersymmetry. In general, the amount of supersymmetry
after dimensional reduction over a Killing vector $\xi$ is given by the number of 
supersymmetry parameters that are invariant along the flow of the vector field. 
This is saying that the Lie derivative of the (parallel) spinor $\epsilon$ along 
$\xi$ should vanish:
\begin{equation}
\mathcal{L}_\xi \epsilon\ =\ \xi^A \mathfrak{D}_A \epsilon 
               + \ft14 \partial_A \xi_B \gamma^{AB} \epsilon \ = \ 0,
\hsp{2cm}
(A,B=0,\dots,10).
\end{equation}
In coordinates adapted to the Killing vector, this implies that the spinor 
does not depend on the compact coordinate: $\partial_z \epsilon=0$.

In our setting, there are 32 parallel eleven-dimensional spinors components $\epsilon$ 
which are constants in natural coordinates. The condition for preserved supersymmetries 
in IIA then reduces to\footnote{Note that this is still a condition on the 
eleven-dimensional spinors $\epsilon$. The condition in terms of the ten-dimensional 
supersymmetry parameters is much more involved. For later convenience it turns out
to be easier to keep explicitly working with the eleven-dimensional parameters for 
the rest of the paper.} 
\begin{equation}
\slashed{B}\epsilon=0\; .
\label{pressusy}
\end{equation}
In other words, the rotation $B$ in our Killing vector should leave 
some subset of the parallel spinors invariant. 
Without loss of generality, $B$ can always be chosen of the following form:
\begin{equation}
B=\left( \begin{array}{cccc}
\rmi \alpha \sigma_2 & 0 & 0 & 0 \\
0 & \rmi \beta \sigma_2 & 0 & 0 \\
0 & 0 & \rmi \gamma \sigma_2& 0 \\
0 & 0 & 0 & 0 
\end{array}\right) , \hsp{1cm} {\rm where} \hsp{1cm}  
\sigma_2=\left( \begin{array}{cc}
0 & -\rmi \\
\rmi & 0
 \end{array}\right)\; ,
\label{B}
\end{equation}
i.e. we can alway choose the rotation $B$ to lie in the Cartan subalgebra of 
$\mathfrak{so}(7)$.\footnote{In order for the Killing vector not to have fixed 
points and the reduced solution to be everywhere non-singular, the rotation $B$ has 
to lie in the $SO(7)$ subgroup of $SO(8)$.} 
Therefore, it should be contained in $\mathfrak{su}(3)$ to preserve 
some of the supersymmetries. In the parametrisation (\ref{B}) this means
\begin{equation}
\alpha+\beta+\gamma=0.
\end{equation}
The number of preserved supersymmetries for different choices of these parameters are 
given in the Table~\ref{table1}, where $Q$ denotes the number of preserved 
supersymmetries and $\mathfrak{g}$ is the subgroup of $\mathfrak{so}(7)$ in 
which the rotation is contained.
\begin{table}[t]
\begin{center}
\begin{tabular}{|c|c|c|}
\hline
&$\mathfrak{g}$&$Q$\\
\hline \hline
$\alpha=\beta=\gamma=0$& \unity&32\\[.2cm]
$\alpha=-\beta  , \; \gamma=0$&$\mathfrak{sp}(1)$&16\\[.2cm]
$\alpha+\beta+\gamma=0$&$\mathfrak{su}(3)$&8\\[.2cm]
$\alpha+\beta+\gamma \neq 0$&$\mathfrak{so}(7)$&0\\[.2cm]
\hline  
\end{tabular}
\end{center}
\caption{\label{table1} 
{\footnotesize The number of supercharges $Q$ of the Type IIA fluxbrane solution 
(\ref{solIIA}) preserved for different choices of $B$. The group
$\mathfrak{g}$ denotes the subgroup of $\mathfrak{so}(7)$ in 
which the rotation is contained.  }
}
\end{table}
Note that if we add an M2- or D2-brane probe, half of the supersymmetries 
will be broken by the probe.

\sect{Hodge-duality on the probes}

In the present Section, we will study this reduction process from the 
viewpoint of an M2/D2-brane probe. We will simplify the theory on the probe 
by gauge fixing the world-volume diffeomorphism invariance and the 
$\kappa$-symmetry and Taylor expanding the membrane (or Born-Infeld) action 
up to terms quadratic in the derivatives of the fields. In that way, we obtain 
non-linear sigma models, and we will point out what happens when a scalar in 
the $\mathcal{N}=8$ theory, corresponding to the M2-brane, is dualised into a 
vector in a theory corresponding to the D2-brane. In particular we expect this 
vector theory not to have $\mathcal{N}=8$ supersymmetry anymore. 

It is well known that a simplification to the level of non-linear sigma models 
could accidentally enhance the number of supersymmetries. A famous example 
being an M2-brane probing orthogonal to an eight-dimensional hyper-K\"ahler 
manifold. The world-volume theory has $\mathcal{N}=3$ although the non-linear 
sigma model has $\mathcal{N}=4$ \cite{Townsend:2002wd}. We will show though, 
that such an enhancement does not appear in our setting. 

\subsection{Probing the bosonic background}
Following the above procedure for an M2-brane probe in eleven-dimensional 
Minkowski space, we find that the resulting non-linear sigma model reads
\begin{equation}
\mathcal{L}_{M2}= -\ft12 \partial_\mu \phi^i \partial^\mu \phi^i -\ft12 
\partial_\mu Z \partial^\mu Z\ .
\label{M2mink}
\end{equation}
We thus trivially find eight non-interacting scalars. The field $Z$ 
corresponds to the position of the brane in the $z$-direction, while 
the $\phi^i$ denotes degrees of freedom in the $x^i$-directions. On the other hand, 
probing the background (\ref{solIIA}) with a $D2$-brane along 
$x^0,\; x^1,\; x^2$, gauge fixing and Taylor expanding yields
\begin{eqnarray}
\mathcal{L}_{D2}=-\half \partial_\mu \chi^i 
\Bigl[ \delta_{ij} \ +\ \Lambda^{-1} B_{ik} \chi^k \ \chi^l B_{lj} \Bigr]
        \partial^\mu \chi^j
\ -\ \ft14 \Lambda^{-1} F_{\mu\nu} F^{\mu\nu} 
\ -\ \ft12 \epsilon^{\mu \nu \kappa}\ \Lambda^{-1} \chi^i B_{ij} 
 \partial_\mu \chi^j\  F_{\nu\kappa},
\label{D2flux}
\end{eqnarray}
where $F_{\mu\nu}= 2\part_{[\mu} A_{\nu]}$ is the field strength of the Born-Infeld 
vector $A_\mu$ and the scalars $\chi^i$ denote the degrees of freedom transverse to 
the D2. $\Lambda$ is again given by  $\Lambda = 1 - \chi^i B_{ij}B_{jk} \chi^k$. 

It is well known that an M2-brane action transforms into a D2-brane Born-Infeld theory 
by dualising the scalar corresponding to the compact direction in the M2-brane action
into the Born-Infeld vector of the D2-brane \cite{Towns}. 
In the same spirit, the theory (\ref{M2mink}) is transformed into (\ref{D2flux}) by 
dualising the coordinate parametrising the flow along the Killing vector $\xi$ 
(\ref{xi}). In order to see this, we should first change to coordinates in the target 
space of \for{M2mink} adapted to $\xi$, meaning that we need the coordinate 
transformation (\ref{chco}). The corresponding field redefinition is obviously
\begin{equation}
\vec{\chi}=e^{-Z B} \vec{\phi},
\label{redefchi}
\end{equation}
and in terms of these new fields, the M2-probe (\ref{M2mink}) reads
\begin{equation}
\mathcal{L}_{M2} =-\ft12 \Lambda\  \partial_\mu Z \partial^\mu Z 
\ - \ \partial_\mu \chi^i B_{ij} \chi^j \ \partial^\mu Z
- \ft12 \partial_\mu \chi^i \partial^\mu \chi^i.
\label{M2twisted}
\end{equation}
In order to dualise $Z$, we substitute $L_\mu = \partial_\mu Z $ and enforce the Bianchi 
identity $\part_{[\mu}L_{\nu]}=0$ by adding  a Lagrange multiplier $A_\mu$ to the 
M2-brane action 
(\ref{M2twisted}):
\begin{equation}
\mathcal{L}=\mathcal{L}_{M2}-\epsilon^{\mu \nu \kappa}A_\mu \partial_\nu 
L_\kappa.
\label{lagrangemult}
\end{equation}
Eliminating $L_\mu$ with its equation of motion yields the dualisation condition for $Z$
\be
\part_\mu Z = -\half \epsilon_{\mu\nu\kappa} \Lambda^{-1} F^{\nu\kappa} 
                  \ - \ \Lambda^{-1} \part_\mu \chi^i B_{ij} \chi^j , 
\label{dualisation}
\ee 
which filled in in the action (\ref{lagrangemult}), yields the D2-brane action 
(\ref{D2flux}).

\subsection{Probing supersymmetric backgrounds} 

We will now extend this procedure to the supersymmetric case and show that 
dualising a scalar into a vector can break supersymmetry, as can be predicted by 
looking at the background.

As Minkowski space is maximally supersymmetric, the theory on the M2-brane 
probe has 16 supersymmetries. This fixes the (non-)linear sigma model of 
(\ref{M2mink}) to the following form ($a=1, ..., 8$):\footnote{Our conventions on fermions
and gamma matrices are explained in the appendix.}
\begin{equation}
{\cal L}_{M2}=  \; \ft12 \partial_\mu \phi^a \partial^\mu \phi^a\ +\ \hat{\bar{\psi}}
\slashed{\partial}\hat{\psi}\; ,
\label{n81}
\end{equation}
while the ${\cal N}=8$ supersymmetry transformation rules are\footnote{Note that this is 
an on-shell algebra.}
\begin{eqnarray}
\delta \phi^a &=& \hat{\bar{\epsilon}}\  \Gamma^a \ \hat{\psi} \; ,\nonumber\\
\delta \hat{\psi}&=&\ft12 \Gamma^a \ \slashed{\partial}\phi^a\ \hat{\epsilon}\; .
\label{n82}
\end{eqnarray}
The R-symmetry algebra is $\mathfrak{so}(8)$ and we used triality to put the 
scalars into the vector representation, while the spinors $\hat \psi$ transform as 
a spinor of both $\mathfrak{so}(8)$ and $\mathfrak{so}(2,1)$ and the parameters for 
supersymmetry $\hat \epsilon$ as a cospinor.

To find the non-linear sigma model corresponding to the D2-brane probe in the 
fluxbrane background, we have to repeat the dualisation procedure in the 
presence of fermions. As we first have to find target space coordinates 
adapted to $\xi$, we have to single out one scalar, just as in (\ref{M2mink}). 
This breaks the R-symmetry to $\mathfrak{so}(7)$. Therefore, we have to choose 
a special base for our $\mathfrak{so}(8)$ gamma matrices
\begin{equation}
\Gamma_*=\sigma_3\otimes \unity_8 , \; \hsp{1cm}
C= \unity_2\otimes \mathcal{C},\;                 \hsp{1cm}
\Gamma^i=\rmi \sigma_2\otimes\tilde{\Gamma}^i,\; \hsp{1cm}
\Gamma^8=-\rmi \sigma_1\otimes \unity_8\; ,
\end{equation}
where $\sigma_1,\sigma_2$ and $\sigma_3$ are the Pauli matrices.
Furthermore we decompose scalars and the spinors as
\begin{equation}
\phi^a = (\phi^i, Z), \hsp{2cm}
\hat{\epsilon}=\left(\begin{array}{c}1\\0\end{array}\right)\otimes\epsilon, \; \hsp{2cm}
\hat{\psi}=\left(\begin{array}{c}0\\1\end{array}\right)\otimes\psi\; ,
\end{equation}
where $\psi$ and $\epsilon$ are Majorana spinor of $\mathfrak{so}(7)$ and 
$\mathfrak{so}(2,1)$. The M2-brane action (\ref{n81}) and the supersymmetry
transformations (\ref{n82}) can then be rewritten as
\begin{equation}
{\cal L}_{M2}^\prime=
-\ft12 \partial_\mu \phi^i \partial^\mu \phi^i
\ -\ \ft12 \partial_\mu Z \partial^\mu Z 
\ - \ \bar{\psi}\slashed{\partial}\psi
\end{equation}
and 
\begin{eqnarray}
\delta \phi^i = \bar{\epsilon}\ \tilde{\Gamma}^i\ \psi\; , \hsp{2cm}
\delta Z = -\rmi \bar{\epsilon}\ \psi\; ,\hsp{2cm} 
\delta \psi=-\ft12 \tilde{\Gamma}^i \ \slashed{\partial}\phi^i\ \epsilon
              \  -\ \ft12  \rmi \slashed{\partial}\phi\  \epsilon\; .
\label{M2transf}
\end{eqnarray}
We subsequently have to do the field redefinition (\ref{redefchi}) on the 
fermions and the $\mathfrak{so}(7)$ gamma matrices as well:
\begin{eqnarray}
\vec{\chi}=e^{-ZB}\vec{\phi},\;   \hsp{1cm}
\theta=e^{-\ft14 Z \slashed{B}} \psi\; , \hsp{1cm}
\gamma^i =  e^{-\tfrac{1}{4} Z \slashed{B}} \left(e^{-ZB}\right)^i {}_j \ 
\tilde{\Gamma}^j \ e^{\ft14 Z \slashed{B}},  \hsp{1cm}
\varepsilon=e^{-\ft14 Z \slashed{B}} \epsilon\; 
\label{redef}
\end{eqnarray}
The resulting superalgebra is given by
\begin{eqnarray}
\delta \chi^i &=& \rmi B_{ij}\  \chi^j\  \bar{\varepsilon}\  \theta
                 \  +\  \bar{\varepsilon}\  \gamma^i \ \theta, 
\nn[.3cm]
\delta Z &=& -\rmi \bar{\varepsilon}\ \theta,
\label{n8susy}\\[.3cm]
\delta \theta &=&  -\ \ft12 \gamma^i \slashed{\partial} \chi^i \ \varepsilon
                 \ - \ \ft12 \rmi \slashed{\partial} Z \ \varepsilon
                 \ +\ \ft14 \rmi \slashed{B} \theta\   \bar{\varepsilon}\ \theta         
                 \  -\ \ft12 \gamma^i B_{ij} \chi^j \slashed{\partial} Z 
                        \varepsilon ,\nonumber
\end{eqnarray}
and the new action Lagrangian 
\begin{equation}\label{newM2}
\mathcal{L}_{M2}=-\ft12 \Lambda\  \partial_\mu Z \partial^\mu Z 
\ - \  \partial_\mu \chi^i B_{ij} \chi^j\  \partial^\mu Z 
\ - \ \ft12 \partial_\mu \chi^i \partial^\mu  \chi^i 
\ - \ \bar{\theta}\slashed{\partial} \theta 
\ -\ \ft14 \bar{\theta} \ \slashed{\partial}Z \ \slashed{B} \ \theta \; ,
\end{equation}
is still ${\cal N} = 8$ supersymmetric, provided that we take in account 
that\footnote{Note that strictly speaking $\varepsilon$ is not a parameter, due to the 
$Z$-dependence.}
\be
\delta_1 \varepsilon_2 = \tfrac{1}{4} \rmi \bar{\varepsilon}_1 \theta \  
                       \slashed{B} \varepsilon_2. 
\ee
To dualise the action, we again write $L_\mu =\partial_\mu Z$, add a Lagrange 
multiplier to enforce the Bianchi identity and subsequently eliminate the 
$L_\mu$ from through its equation of motion, which is the duality condition
\begin{equation}
\partial_\mu Z=  -\ft12 \epsilon_{\mu \nu \kappa} \Lambda^{-1}\ F^{\nu \kappa}
- \Lambda^{-1} \partial_\mu \chi^i B_{ij} \chi^j 
- \ft14  \Lambda^{-1} \bar{\theta}\ \tau_\mu \slashed{B}\ \theta
 \label{dual}\; .
\end{equation}
The resulting dualised action now reads
\begin{eqnarray}
\mathcal{L}_{D2}&=&
-\ \ft12 \partial_\mu \chi^i \partial^\mu \chi^i 
\ -\ \ft12 \Lambda^{-1} \partial_\mu \chi^i B_{ij} \chi^j\  \chi^k B_{kl} \partial^\mu \chi^l
\ -\ \ft14 \Lambda^{-1} F^2 
\ - \ \ft12 \epsilon^{\mu \nu \kappa}\Lambda^{-1} \chi^i B_{ij} 
             \partial_\mu \chi^j F_{\nu\kappa} 
\nonumber\\ [.3cm]
&&-\ \bar{\theta} \slashed{\partial} \theta
\ -\ \ft1{8} \Lambda^{-1} \bar{\theta} \slashed{F} \slashed{B} \theta 
\ - \ \ft1{4}\Lambda^{-1}  B_{ij} \chi^i \ \bar{\theta}\  \slashed{\partial} \chi^j
               \slashed{B}\  \theta 
\ +\ \ft1{32} \Lambda^{-1} \bar{\theta}\ \tau_\mu \slashed{B} \ \theta \  
                \bar{\theta}\ \tau^\mu \slashed{B}\ \theta\; .
\label{D2dualaction}
\end{eqnarray}
Since the dualised action has become independent of $Z$, we interpret it as the action of 
an D2-brane probe in the background of the fluxbrane (\ref{solIIA}). The amount of 
supersymmetry preserved after the dualisation depends on the supersymmetry preserved by 
the background (\ref{solIIA}) and is given by the rotation $B$ in the Killing vector. 
In analogy with the action, we demand the supersymmetry rules of (\ref{D2dualaction}) 
to be related to (\ref{n8susy}). Actually, we ask that the transformation rules for 
$\chi^i$ and $\theta$ remain the same, but substituting the $\partial_\mu Z$-factors by 
their duality condition (\ref{dual}). Consistency requires however that all 
$Z$-dependence should drop out of the transformation rules, in particular also the 
$Z$-dependence in the supersymmetry parameter $\varepsilon$. We therefore also have 
to impose the following condition on $\varepsilon$:
\be
\part_Z \ \varepsilon = 0  
\hsp{1cm} \Longrightarrow \hsp{1cm}
\slashed{B}\ \epsilon=0, \hsp{.5cm}  \varepsilon = \epsilon.
\label{pressusy2}
\end{equation}
We thus see that only those components of $\epsilon$ survive that are annihilated by 
$\slashed{B}$, which reduces the amount of supersymmetry. Note that this is precisely the
condition (\ref{pressusy}) we found for the background (see Section 4 for a discussion 
here on).
 
Finally, we still have to find the transformation of the vector, with the condition that
the supersymmetry algebra should close and the action (\ref{D2dualaction}) should be 
invariant. This can be done by asking that the dualisation condition (\ref{dual}) is invariant under supersymmetry (using the equation of motion for the fermions). Indeed we find that for the supersymmetry transformation rules 
\begin{eqnarray}
\delta \chi^i &=& \rmi B_{ij} \chi^j \ \bar{\epsilon}\theta 
                 \ +\  \bar{\epsilon}\ \gamma^i\ \theta \; , 
\nonumber\\[.3cm]
\delta A_\mu &=& \rmi \bar{\epsilon}\ \tau_\mu \theta 
                 \ -\ \bar{\epsilon}\ \tau_\mu \gamma^i \ \theta\ B_{ij} \chi^j\; , 
\label{vectorsusy}
\\ [.3cm]
\delta \theta&=&-\ft12 \gamma^i \slashed{\partial} \chi^i \epsilon
\ +\ \ft1{2} \Lambda^{-1} \gamma^i\slashed{\partial} \chi^k \epsilon 
             \ B_{ij} \chi^j \ B_{kl} \chi^l 
\ +\  \ft1{2} \rmi \Lambda^{-1} \slashed{\partial} \chi^i \epsilon\  B_{ij} \chi^j
-\ \ft1{4} \Lambda^{-1} \gamma^i \slashed{F} \epsilon\  B_{ij} \chi^j 
\nonumber \\[.3cm]
&& - \ \ft1{4} \rmi \Lambda^{-1} \slashed{F} \epsilon
\ +\ \ft14 \rmi \slashed{B} \theta \ \bar{\epsilon} \theta 
\ + \ \ft1{8}\Lambda^{-1} \gamma^i \tau^\mu \epsilon 
            \  \bar{\theta} \tau_\mu \slashed{B} \theta \ B_{ij} \chi^j  
\ +\  \ft1{8} \rmi \Lambda^{-1} \tau^\mu \epsilon\ 
             \bar{\theta} \tau_\mu \slashed{B} \theta\; , 
\nonumber
\end{eqnarray}
the supersymmetry algebra closes on-shell and the D2-brane probe action 
(\ref{D2dualaction}) is invariant up to terms proportional to $\slashed{B} \epsilon$, 
which are set to zero. The total number $Q$ of supercharges preserved is thus determined 
by $B$ and can be found in Table \ref{table2}.
\begin{table}[t]
\begin{center}
\begin{tabular}{|c|c|c|}
\hline
&$\mathfrak{g}$&$Q$\\
\hline \hline
$\alpha=\beta=\gamma=0$&\unity&16\\[.2cm]
$\alpha=-\beta  , \; \gamma=0$&$\mathfrak{sp}(1)$&8\\[.2cm]
$\alpha+\beta+\gamma=0$&$\mathfrak{su}(3)$&4\\[.2cm]
$\alpha+\beta+\gamma \neq 0$&$\mathfrak{so}(7)$&0\\[.2cm]
\hline  
\end{tabular}
\end{center}
\caption{\label{table2} 
{\footnotesize The number of supercharges $Q$ of the D2-brane probe
(\ref{D2dualaction}) preserved for different choices of $B$. The group
$\mathfrak{g}$ denotes the subgroup of $\mathfrak{so}(7)$ in 
which the rotation is contained. }}
\end{table}

In summary, from a $D=3$, ${\cal N} = 8$ theory with eight scalars, we have constructed 
via 
the dualisation of one of the scalars into a vector, the Lagrangian (\ref{D2dualaction}),
which is still invariant under the supersymmetry transformations (\ref{vectorsusy}). The 
obtained theory with seven scalars and one vector is however in general no longer 
${\cal N} = 8$ supersymmetric, but some of the supersymmetries get broken, due to the 
dependence of the supersymmetry parameter on the dualised coordinate.

\sect{General considerations}
In the previous Section, we have shown that it is possible to dualise a 
scalar in the $\mathcal{N}=8$ multiplet in such a way that the dualised 
theory has less supersymmetry. This was to be expected as we were probing 
with a non-linear sigma model a supergravity background which was losing 
supersymmetry while dimensionally reducing. We will now show that this breaking 
by dualisation is quite general.

The theory (\ref{n81})-(\ref{n82}) is maximally supersymmetric and realises 
the extended super-Poincar\'e algebra. We now look for an extra $\mathfrak{u}(1)$ or 
$\mathbb{R}$ isometry $G$, which acts as 
\begin{eqnarray}
\delta_G \phi^a=\lambda \ \xi^a\; , \hsp{2cm}
\delta_G \psi=\lambda \ t \cdot \psi, 
\label{Gpsi}
\end{eqnarray}
where the parameter $\lambda$ is a constant, $\xi^a$ is a Killing vector of 
target space,\footnote{I.e. it leaves the action (\ref{n81}) invariant.}  and $t$ is the 
representation of this isometry on the fermions, whose explicit form will be determined 
shortly, but that consists of an even number of gamma matrices to preserve the 
$\mathfrak{so}(8)$ chirality of the fermions under the transformations $G$: 
$t= t_0 + t_{ab} \Gamma^{ab} + t_{abcd} \Gamma^{abcd}$. We would like to find for which 
$\xi^a$ and $t$ (part of the) supersymmetry is realised. The only non-trivial condition 
will come from the commutator of the isometry transformations with the supersymmetry 
(\ref{n82}) .  

In the most general case, the commutator of an external symmetry and a soft
supersymmetry algebra can give a new supersymmetry transformation with a field 
dependent parameter $\eta$:
\be
[\delta_G, \delta_Q] = \delta_Q (\eta). 
\ee
When we try to realise this for the isometry (\ref{Gpsi}) and the supersymmetry 
transformations (\ref{n82}), we notice that such a realisation is trivial: the only 
transformations (\ref{n82}) that preserve all the supersymmetry are the 
translations 
\bea
\delta_G \phi^a = C^a, \hsp{2cm}
\delta_G \psi = 0, 
\label{translaties} 
\eea
($C^a$ is a constant vector)
and they necessarily commute with the supersymmetry transformations (\ref{n82}).
This is of course the well known case of standard reduction and dualisation in an isometry 
direction that preserves all supersymmetry.

It is however possible to look for more general cases, where only a subgroup of the 
Poincar\'e algebra commutes with the isometry transformations (\ref{Gpsi}), which will in 
general break supersymmetry. When we try to close the commutator of (\ref{Gpsi}) and 
the supersymmetry (\ref{n82}) on the scalars and the fermions, we find the conditions
\bea 
&&[\delta_G, \delta_Q] \phi^a \ =\ - \lambda \ \bar{\psi}\ \Bigl[
 \Gamma^b \part_b \xi^a -  (t_{(0)} - t_{(2)} + t_{(4)} ) \Gamma^a  \Bigr] \epsilon = 0 ,
\label{t1} \\ [.3cm]
&&[\delta_G, \delta_Q] \psi \ =\   -\half \lambda \Bigl[
       \Gamma^b \part_b \xi^a + ( t_{(0)} + t_{(2)} + t_{(4)} ) \Gamma^a \Bigr] 
\slashed{\part} \phi^a \epsilon = 0, 
\label{t2}
\eea
where we took $\part_a t=0$ because it is the only commutator term in (\ref{t2}) quadratic in the 
fermions. Equating (\ref{t1}) and (\ref{t2}) then yields
\bea
&& t_{(0)} = t_{abcd} = 0, \nn [.3cm]
&& \Bigl[ \Gamma^b \part_b \xi^a 
        + t_{bc} \Bigl( \Gamma^a \Gamma^{bc} - 4 \delta^{ab} \Gamma^{c} \Bigr)\Bigr] 
                 \epsilon = 0 .
\label{conditon}
\eea
The latter condition has a simple solution of the form 
\begin{equation}
\label{sol}
t_{ab}=-\ft14 \partial_{[a} \xi_{b]}, 
\hsp{2cm}
\Gamma^{ab} t_{ab} \ \epsilon = 0.
\end{equation}
Hence, isometry transformations of the form (\ref{Gpsi}) will commute with the 
supersymmetry algebra and preserve that part of the supersymmetry for which 
$\slashed{t} \epsilon = 0$. Note that this coincides with the condition (\ref{pressusy}) 
and (\ref{pressusy2}). Dualisation of a scalar along the Killing vector $\xi$ is precisely 
what we have done in Section 3.  

It is now obvious that these considerations are quite general. In dimensional reduction, 
the (super)-isometry algebra of the reduced solution is that part of the isometry algebra 
of the higher-dimensional field configuration that commutes with the Killing vector used 
during the reduction. Therefore, if we look at the theory on the probes, the part of the 
supersymmetry algebra commuting with the Killing vector will in general be realised after 
the dualisation.
 
\sect{Conclusion}

We have shown how the world-volume supersymmetry can be broken in the Hodge-dualisation
involved in transforming the world-volume effective action of an M2-brane into a D2-brane, 
at the level of the non-linear sigma model. In particular, we have looked at an M2-brane 
probe in eleven-dimensional Minkowski space and reduced this over a non-trivial Killing 
vector (involving both translations and rotations) to a D2-brane probe in a fluxbrane 
background. Dualising one of the 8 scalars in the $D=3$, ${\cal N} = 8$ theory living on 
the world-volume of the M2-brane, we obtain a $D=3$ theory with seven scalars and one 
vector on the D2-brane, that has less then ${\cal N} = 8$ supersymmetry. At the level of 
the background, the supersymmetry is broken through the dependence of the Killing spinor 
on the Killing vector. At the level of the M2-brane world-volume theory, this translates
into the fact that the algebra has explicit dependence on $Z$ (not only via its 
derivatives). Consistency of the Kaluza-Klein picture requires this dependence to vanish 
in the D2-brane theory, such that only the components of the supersymmetry parameter 
survive that are invariant under the action of the Killing vector. This technique is 
similar to the well-known feature of supersymmetry breaking in dimensional reduction in 
supergravity solutions, as pointed out in \cite{Bakas, BKO}, but to our knowledge it is 
the first time that it is performed at the level of the non-linear sigma models of brane 
effective actions.

As in the case of dimensional reduction of supergravity solutions, this supersymmetry 
breaking is a consequence of the fact that we are limiting ourselves to the level of 
the non-linear sigma model. If we would take in account the full Born-Infeld effective 
actions of the M2- and the D2-brane, the spinor components that do depend on the 
compactified and dualised coordinate would still enter in the game as higher-order 
Kaluza-Klein modes and supersymmetry would the completely preserved, analogously to what 
happens at at the level of the string world-sheet in the case of T-duality \cite{BS, 
Hassan, AAB, Sfetsos}. However, since the non-linear sigma-model is a consistent 
truncation of the full membrane effective action, our results hold in this range of 
validity.   

It is well-known that $D=3$ theories with 8 scalars and ${\cal N} = 4$ (${\cal N} = 2$) 
supersymmetry have hyper-K\"ahler (Calabi-Yau) target manifolds. An interesting question
that arises then is whether the target manifolds that appear in the ${\cal N} = 4$ 
(${\cal N} = 2$) theories (\ref{D2dualaction}) with seven scalars and one vector are in 
some sense related to these hyper-K\"ahler (Calabi-Yau) target manifolds. In other words,
whether it would be possible to dualise the vector back into a scalar, but without 
enhancing the supersymmetry back to ${\cal N} = 8$.  An obvious example is given by 
the action (\ref{newM2}), keeping the condition that $\slashed{B} \epsilon = 0$, for the 
different choices of $B$ as given in Table \ref{table2}. Of course, the resulting theory 
can then easily be extended to the original ${\cal N} = 8$ theory. A less trivial 
solution would be to add some (${\cal N} = 8$) supersymmetry breaking potential to the 
M2-brane action (\ref{newM2}) and add extra terms to the supersymmetry variations in 
(\ref{M2transf}), such that the algebra closes and the action is invariant up to terms 
proportional to  $\slashed{B} \epsilon$. It turns out however, that modifications of
the M2-brane action and the supersymmetry transformations become very involved and lead 
inevitably to modifications in the D2-brane side as well, such that it will no longer be 
clear what the relation is with the obtained action (\ref{D2dualaction}). 

Finally we would like to point out that our results can be easily extended from
probes in Minkowski space to probes in any eleven-dimensional supersymmetric background 
admitting an isometry, leading to for instance probes in spaces describing supersymmetric
configurations of D-branes or NS-branes embedded in fluxbrane backgrounds \cite{FS2, FS3}.

\vspace{1.4cm}
\noindent
{\bf Acknowledgements}
\vsp{.3cm}

\noindent
The authors are grateful to Toine van Proeyen for the useful discussions.  
This work has been made possible with the aid of the F.W.O.-Vlaanderen, to
which J.G.~is associated as an Aspirant-F.W.O. and B.J.~as a Post-doctoral 
Fellow. This work was also partially supported by the F.W.O.-project 
G0193.00N, the European Commission R.T.N.-program HPRN-CT-2000-00131 
and by the Federal Office for Scientific, Technical and Cultural Affairs 
through the Interuniversity Attraction Pole P5/27.

\newpage
\appendix
\renewcommand{\theequation}{\Alph{section}.\arabic{equation}}

\section{Conventions}
We use the convention of \cite{proeyen}.
The $x^{10}$ direction is denoted by $z$. We anti-symmetrise with unit coefficient.
and always use a mostly plus metric. The covariant derivative on a spinor reads
\begin{equation}
\mathfrak{D}_\mu \epsilon=\partial_\mu \epsilon+\tfrac{1}{4} 
\slashed{\omega}_{\mu} \epsilon\ .
\end{equation}

\subsection{$SO(2,1)$ spinors }
The $SO(2,1)$ spinors are Majorana (i.e. 
$\bar{\lambda}\ =\ \lambda^T \mathcal{C}
              \ =\ \alpha^{-1} \lambda^\dagger \gamma_0$, 
with $\mathcal{C}$ charge conjugation and $\alpha$ a unimodular constant). 
We take the spinors to be Grassman valued.
The Clifford algebra is given by
\begin{equation}
\{ \tau^\mu, \tau^\nu \}=2\eta^{\mu \nu}
\end{equation}
with $\tau_2=-\tau_0 \tau_1$. The gamma-matrices can be represented by 
Pauli-matrices as
\begin{equation}
\tau_0=i \sigma_1, \hsp{2cm}
\tau_1=\sigma_2, \hsp{2cm}
\tau_2=\sigma_3.
\end{equation}
Hodge duality imposes the following relations between the gamma matrices
\bea
&&\tau_{\mnr}=-\epsilon_{\mnr},  \hsp{2cm} 
\tau_\mu =\ft12 \epsilon_{\mnr} \tau^{\nu\rho}, \nonumber\\[.3cm]
&&\tau_{\mn}=-\epsilon_{\mnr} \tau^\rho,    \hsp{1.8cm}   
1=\ft16 \epsilon_{\mnr} \tau^{\mnr}.
\end{eqnarray}
For $\bar{\lambda} \tau^{(n)} \chi=t_{(n)} \bar{\chi} \tau^{(n)} \lambda$, 
we find the following transposition properties
\begin{center}
\begin{tabular}{|c||c|c|c|c|} \hline
$n$&0&1&2&3 \\ \hline
$t_{(n)}$ & +&-&-&+ \\ \hline
\end{tabular}
\end{center}
and the $SO(2,1)$ Fierz identity is given by
\begin{equation}
\chi \bar{\lambda}=-\ft12\Bigl(\bar{\lambda}\chi + \bar{\lambda}\tau_\mu \chi \tau^\mu \Bigr).
\end{equation}

\subsection{$SO(8)$ spinors}
The SO(8) spinors are Majorana-Weyl and we take them to be Grassman even.
The chirality matrix $\Gamma_*$ is defined as $\Gamma_*=\prod_{a=1}^8 \Gamma^a$  and 
$C$ is charge conjugation. 
For $\bar{\lambda} \Gamma^{(n)} \chi=t_{(n)} \bar{\chi} \Gamma^{(n)} \lambda$, 
we find the following transposition properties
\begin{center}
\begin{tabular}{|c||c|c|c|c|}\hline
$n$ & 0,4,8 & 1,5 & 2,6 & 3,7 \\ \hline
$t_{(n)}$ & + & + & - & - \\ \hline
\end{tabular}
\end{center}
In the case that three spinors have the same chirality, we have
the following Fierz-identity
\begin{equation}
\eta \bar{\lambda} \chi=
\tfrac{1}{16}\Bigl (2\bar{\lambda} \eta \chi
               \ -\ \bar{\lambda}\Gamma_{ab} \eta  \Gamma^{ab}\chi
               \ +\ \tfrac{1}{4!}\bar{\lambda} \Gamma_{abcd}\eta 
                    \Gamma^{abcd}\chi \Bigr).
\end{equation}

\subsection{$SO(7)$ spinors}
These spinors are Majorana and Grassman even. Gamma matrices are $\tilde{\Gamma}^i$ and
$\gamma^i$, 
charge conjugation is $\mathcal{C}$.
The $SO(7)$ gamma matrices satisfy the following Hodge duality relations
\begin{equation}
\gamma_{i_1 \dots i_n}=-\ft1{(7-n)!} \rmi \epsilon_{i_1 \dots i_7}
             \gamma^{i_7 \dots i_{7-n}}.
\end{equation}
For $\bar{\lambda} \gamma^{(n)} \chi=t_{(n)} \bar{\chi} \gamma^{(n)} \lambda$, 
we find the following transposition properties
\begin{center}
\begin{tabular}{|c||c|c|c|c|} \hline
$n$ & 0,4 & 1,5 & 2,6 & 3,7 \\ \hline
$t_{(n)}$ & + & - & - & + \\ \hline
\end{tabular}
\end{center}
and the Fierz identity is given by
\begin{equation}
\lambda \bar{\chi}=\ft18\Bigl(\bar{\chi}\lambda
                \ + \ \bar{\chi}\gamma^i\lambda\gamma_i
                \ -\ \ft12\bar{\chi}\gamma^{ij}\lambda\gamma_{ij}
                 \ -\  \ft16\bar{\chi}\gamma^{ijk}\lambda\gamma_{ijk} \Bigr).
\end{equation}


\end{document}